\documentclass[%
 reprint,
superscriptaddress,
 amsmath,amssymb,
 aps,
]{revtex4-1}

\usepackage{tabularx}
\usepackage{booktabs}
\usepackage{upgreek}
\usepackage{nicefrac}
\usepackage{hyperref}
\usepackage{natbib}
\usepackage{xcolor}
\usepackage{graphicx}% Include figure files
\usepackage{dcolumn}% Align table columns on decimal point
\usepackage{bm}% bold math
\usepackage{orcidlink}
\usepackage{upgreek}

\begin{document}

\preprint{APS/123-QED}

\title{Nuclear charge radii of silicon isotopes}% Force line breaks with \\

\author{Kristian K\"onig\orcidlink{0000-0001-9415-3208}}
 \email{kkoenig@ikp.tu-darmstadt.de}
\affiliation{Facility for Rare Isotope Beams, Michigan State University, East Lansing 48824, USA}
\affiliation{Technische Universtität Darmstadt, 64289 Darmstadt, Germany}

\author{Julian C. Berengut}
\affiliation{School of Physics, University of New South Wales, NSW 2052, Australia}

\author{Anastasia Borschevsky}
\affiliation{University of Groningen, 9747 Groningen, Netherlands}

\author{Alex Brinson}
\affiliation{Massachusetts Institute of Technology, Department of Physics, Cambridge, MA 02139}

\author{B. Alex Brown\orcidlink{0000-0002-6111-1906}}
\affiliation{Facility for Rare Isotope Beams, Michigan State University, East Lansing 48824, USA}
\affiliation{Department of Astronomy and Physics, Michigan State University, East Lansing 48824, USA}

\author{Adam Dockery\orcidlink{0000-0002-3066-978X}}
\affiliation{Facility for Rare Isotope Beams, Michigan State University, East Lansing 48824, USA}
\affiliation{Department of Astronomy and Physics, Michigan State University, East Lansing 48824, USA}

\author{Serdar Elhatisari\orcidlink{0000-0002-7951-1991}}
\affiliation{%
Faculty of Natural Sciences and Engineering, Gaziantep Islam Science and Technology University, Gaziantep 27010, Turkey
}%
\affiliation{%
Helmholtz-Institut für Strahlen- und Kernphysik and Bethe Center for Theoretical Physics, Universit\"at Bonn, D-53115 Bonn, Germany
}%

\author{Ephraim Eliav}
\affiliation{School of Chemistry, Tel Aviv University, 69978 Tel Aviv, Israel}

\author{Ronald F. Garcia Ruiz}
\email{rgarciar@mit.edu}
\affiliation{Massachusetts Institute of Technology, Department of Physics, Cambridge, MA 02139}

\author{Jason D. Holt\orcidlink{0000-0003-4833-7959}}
\affiliation{TRIUMF, Vancouver, British Colombia, V6T 2A3 Canada}
\affiliation{Department of Physics, McGill University, Montr\'eal, QC H3A 2T8, Canada}

\author{Bai-Shan Hu\orcidlink{0000-0001-8071-158X}}
\affiliation{National Center for Computational Sciences, Oak Ridge National Laboratory, Oak Ridge, Tennessee 37831, USA}
\affiliation{Physics Division, Oak Ridge National Laboratory, Oak Ridge, Tennessee 37831, USA}

\author{Jonas Karthein\orcidlink{0000-0002-4306-9708}}
\affiliation{Massachusetts Institute of Technology, Department of Physics, Cambridge, MA 02139}

\author{Dean Lee\orcidlink{0000-0002-3630-567X}}
\affiliation{Facility for Rare Isotope Beams, Michigan State University, East Lansing 48824, USA}
\affiliation{Department of Astronomy and Physics, Michigan State University, East Lansing 48824, USA}%

\author{Yuan-Zhuo Ma\orcidlink{0000-0002-0892-4457}}
\affiliation{Facility for Rare Isotope Beams, Michigan State University, East Lansing 48824, USA}
\affiliation{Department of Astronomy and Physics, Michigan State University, East Lansing 48824, USA}%

\author{Ulf-G. Mei{\ss}ner\orcidlink{0000-0003-1254-442X}}
\affiliation{%
Helmholtz-Institut für Strahlen- und Kernphysik and Bethe Center for Theoretical Physics, Universit\"at Bonn, D-53115 Bonn, Germany
}%

\author{Kei Minamisono\orcidlink{0000-0003-2315-5032}}
\email{minamiso@frib.msu.edu}
\affiliation{Facility for Rare Isotope Beams, Michigan State University, East Lansing 48824, USA}
\affiliation{Department of Astronomy and Physics, Michigan State University, East Lansing 48824, USA}

\author{Alexander~V.~Oleynichenko\orcidlink{0000-0002-8722-0705}}
\affiliation{Petersburg Nuclear Physics Institute named by B.~P.~Konstantinov of NRC ``Kurchatov Institute'', Gatchina 188300, Russia}

\author{Skyy Pineda}
\affiliation{Facility for Rare Isotope Beams, Michigan State University, East Lansing 48824, USA}
\affiliation{Department of Chemistry, Michigan State University, East Lansing 48824, USA}

\author{Sergey D. Prosnyak\orcidlink{0000-0003-0335-4747}}
\affiliation{Petersburg Nuclear Physics Institute named by B.~P.~Konstantinov of NRC ``Kurchatov Institute'', Gatchina 188300, Russia}
\affiliation{Saint Petersburg State University, 7/9 Universitetskaya nab., St. Petersburg, 199034 Russia}

\author{Marten L. Reitsma}
\affiliation{University of Groningen, 9747 Groningen, Netherlands}

\author{Leonid V. Skripnikov\orcidlink{0000-0002-2062-684X}}
\affiliation{Petersburg Nuclear Physics Institute named by B.~P.~Konstantinov of NRC ``Kurchatov Institute'', Gatchina 188300, Russia}
\affiliation{Saint Petersburg State University, 7/9 Universitetskaya nab., St. Petersburg, 199034 Russia}

\author{Adam Vernon\orcidlink{ 0000-0001-8130-0109}}
\affiliation{Massachusetts Institute of Technology, Department of Physics, Cambridge, MA 02139}

%\author{Shane Wilkins}
%\affiliation{Massachusetts Institute of Technology, Department of Physics, Cambridge, MA 02139}

\author{Andr\'ei Zaitsevskii\orcidlink{0000-0002-5347-2429}}
\affiliation{Petersburg Nuclear Physics Institute named by B.~P.~Konstantinov of NRC ``Kurchatov Institute'', Gatchina 188300, Russia}
\affiliation{Department of Chemistry, M. V. Lomonosov Moscow State University, Leninskie gory 1/3, Moscow, 119991 Russia}

\date{\today}% It is always \today, today,

\begin{abstract}
The nuclear charge radius of $^{32}$Si was determined using collinear laser spectroscopy. The experimental result was confronted with {\it ab initio} nuclear lattice effective field theory, valence-space in-medium similarity renormalization group, and mean field calculations, highlighting important achievements and challenges of modern many-body methods.
The charge radius of $^{32}$Si completes the radii of the mirror pair $^{32}$Ar - $^{32}$Si, whose difference was correlated to the slope $L$ of the symmetry energy in the nuclear equation of state. Our result suggests $L \leq 60$\,MeV, which agrees with complementary observables.%, such as the parity-violating asymmetry, nuclear reactions, and gravitational wave observations.
\end{abstract}

\maketitle

{\it Introduction.}
\label{sec:Introduction}
Recent advances in many-body methods, the continuous increase in computing power, and the development of inter-nucleon potentials derived from Chiral Effective Field Theory (Chiral-EFT), are leading up to a new era of precision nuclear theory calculations with quantifiable uncertainties \cite{Epelbaum:2008ga,Her20,Bai22}. Besides the description of diverse nuclear properties, even extremely neutron-rich matter, such as neutron stars, can now be addressed \cite{Cap20,Drischler.2020}. 

The properties of neutron stars are governed by the nuclear equation of state (EOS) and affect, for instance, the forms of gravitational waves from a binary neutron star merger \cite{Abbot.2017} or the character of super heavy nuclei \cite{Nazarewicz.2018}. However, despite the broad experimental efforts, the form of the EOS, especially the slope $L$ in the symmetry energy, could only be constrained to a limited range \cite{Roca-Maza.2018, Horowitz.2014, Vinas.2014} insufficient for precise model predictions. An alternative approach to constrain $L$ based on the concept of charge symmetry of the nuclear interaction was suggested recently. It uses the differences of charge radii of a pair of mirror nuclei as a proxy for the neutron-skin thickness \cite{Wang.2013, Brown.2017, Yang.2018}, for which the correlation on $L$ was recently discussed in \cite{Reinhard.2022b,Brown.2022,Huang.2023,An.2023,Bano.2023}. Enhanced sensitivity is expected thanks to possibly large isospin asymmetry if one of the mirror nuclei is a radioactive nucleus \cite{Piekarewicz.2011, Brown.2017}, hence contrasting most previous studies on stable nuclei due to technical reasons.

Despite the compelling progress in nuclear theory, significant long-standing challenges persist in our understanding of nuclei \cite{Elhatisari:2022qfr}. For instance, obtaining a simultaneous description of the binding energy and nuclear charge radii has proven to be a major challenge \cite{Eks15,Gar16,deGroote.2020,Elhatisari:2022qfr}. Moreover, it is still unclear if effective theories constrained to a finite number of nuclei can provide reliable calculations of infinite nuclear matter at supersaturation density \cite{Tsa12,Hagen.2016, Brown.2017,Drischler.2020}. Therefore, precision measurements of charge radii for nuclei with large proton-to-neutron asymmetries are critical in guiding the progress of nuclear theory and the description of nuclear matter.  

The investigation of nuclear charge radii of Si isotopes ($Z=14$), in particular, highlights several questions of great interest. The charge radius of $^{32}$Si, measured in this work, sets a new constraint on $L$ when combined with data of its mirror partner $^{32}$Ar \cite{Klein.1996}. Furthermore, nuclear charge radii of Si isotopes play a critical role in studies of the appearance or disappearance of nuclear magic numbers \cite{Piekarewicz.2007,Fridmann.2005} and the emergence of exotic nuclear shapes, e.g., bubble nuclei \cite{Mutschler.2016, Duguet.2017}. From the theoretical side, recent progress was made in calculating these properties by several many-body methods \cite{Duguet.2017,Elhatisari:2022qfr}. 

However, previous to our work, only measurements for the nuclear charge radii of stable silicon isotopes were available \cite{Angeli.2013,Yang23}. This is partly because nuclear charge radii measurements of short-lived Si isotopes pose major challenges in production and extraction from the thick targets of Isotope Separator Online (ISOL) facilities. Moreover, silicon is a highly reactive element and likely to form molecular compounds unsuitable for laser spectroscopy experiments. Here, we present measurements of the differential charge radius of $^{32}$Si obtained from collinear laser spectroscopy of $^{28,29,30,32}$Si isotopes after the molecular break-up of SiO molecules. The experiment was performed at the BECOLA setup at the Facility for Rare Isotope Beams (FRIB).

{\it Experiment.}
\label{sec:Setup}
The stable $^{28,29,30}$Si isotopes used as a reference for the isotope shift measurements were produced in a Penning-ionization gauge (PIG) ion source \cite{Ryder.2015} with cathodes of natural silicon.  
The radioactive $^{32}$Si beam was generated with an oven-ion source (OIS) of the batch-mode ion source system (BMIS) \cite{Sumithrarachchi.2023}. The OIS is a CERN/ISOLDE target ion source unit, in which Si powder containing $^{32}$Si was installed and heated up to 1500$^\circ$C to surface ionize Si atoms. The beam, generated $\sim$20\,m upstream of BECOLA, first went through a dipole magnet for mass selection. Due to a 1000:1 contamination of $^{32}$S relative to $^{32}$Si at mass 32, a mass-48 beam was selected instead. This mass component was the most populated from the BMIS and contained mostly singly-charged $^{32}$Si$^{16}$O. 

At BECOLA \cite{Minamisono.2013,Rossi.2014}, the ions were first fed into a helium-gas-filled radio-frequency quadrupole (RFQ) ion trap \cite{Barquest.2017} floated at a potential of 29813\,V. The helium-buffer gas pressure was set to 120 mTorr. Since the beam energy from the ion sources was 30\,keV, the injection energy into the RFQ was about 190\,eV. This injection energy was sufficient to break the SiO molecules by collisions with the helium buffer gas to be left with bare singly charged Si ions required for laser spectroscopy. At a 100\,V lower injection energy, no laser spectroscopy signal was observed due to a low dissociation efficiency. At a 100 V higher injection energy, the stopping efficiency in the RFQ was decreased, and only a weak resonance signal was observed. The resulting bare singly charged Si ions were cooled by collisions with the He buffer gas and then extracted as a continuous beam. %The radio frequency of the RFQ for the transverse confinement was set so that the dissociated oxygen ions were unstable in the trap and rejected.
%Bunching the ion beam was not employed in the present measurement to avoid systematic shifts of the trapping potential due to the space charge caused by different beam contamination between the stable Si beams from the PIG source and the radioactive $^{32}$Si from the BMIS.

Since Si$^+$ ions are not accessible by laser spectroscopy due to the lack of transitions in the optical regime, the ions were first neutralized with Na vapor inside the charge-exchange cell (CEC) \cite{Klose.2012}. The CEC was heated to 410$^\circ$C, leading to a 50\,\% neutralization efficiency of the incoming ion beam. During the charge exchange process, many electronic states are populated and redistributed through spontaneous decays. In this cascade decay, low-lying meta-stable states tend to be populated. One of these meta-stable states ($3s^2 3p^2\;^1\mathrm{S}_0$, 15,394.370 cm$^{-1}$ \cite{NIST}) was used as the lower state for the laser spectroscopy. Based on a simulation \cite{Ryder.2015}, about 2\% of the total population reached this $^1\mathrm{S}_0$ state at the time ions arrived in the fluorescence detection region (FDR) installed 70\,cm downstream of the CEC.

The atoms in the $3s^2 3p^2\;^1\mathrm{S}_0$ state were excited with laser light at 391\,nm to the $3s^2 3p 4s\;^1\mathrm{P}_1^o$ state at 40,991.884 cm$^{-1}$.
With a probability of 93\,\%, the excited electrons decay at 288\,nm to the $3s^2 3p^2\;^1\mathrm{D}_2$ state at 6,298.850 cm$^{-1}$ \cite{NIST}, which allowed us to perform laser-background-free spectroscopy by eliminating the scattered 391-nm light. 
Therefore, we used a Hamamatsu H11870-09 photo-multiplier tube (PMT) with a quantum efficiency of $\approx 7$\,\% at 288\,nm but a four orders of magnitude lower sensitivity at 391\,nm. Additionally, we placed an absorption filter (Hoya U340) in front of the PMT, which transmits more than 50\,\% of the UV light but absorbs 99.8\,\% of the scattered 391\,nm laser light.
The fluorescence light was collected with an elliptical mirror with MIRO coating from ALANOD, which is highly reflective in the deep UV light. The PMT was placed outside the vacuum chamber at the second focal point. 
To achieve resonance between atom and laser frequency, Doppler tuning was applied. The atom velocity was altered by applying a small scanning potential difference of $<50$\,V to the CEC, causing Doppler shifts of the transition frequency, while the laser frequency was kept constant.
The employed laser was a continuous-wave Ti-sapphire laser (Matisse TS, Sirah Lasertechnik) operated at 782\,nm and pumped by a frequency-doubled Nd-YAG solid-state laser (Millennia eV, Spectra Physics). The 782-nm light was guided to a cavity-based frequency doubler (Wavetrain, Spectra Physics), creating the 391-nm light. This light was transported via an optical fiber to the beamline and irradiated in collinear geometry. In front of the optical fiber, an acousto-optical modulator (AOM) was used to block the laser beam. Spectroscopy was performed with a laser power of 4\,mW and a laser-beam diameter of 2\,mm at the interaction region.
The short-term frequency stabilization was realized via side-of-fringe locking to a reference cavity. For long-term stabilization, the cavity length was controlled by feedback from a wavelength meter (WSU30, HighFinesse) calibrated every minute to a helium-neon laser (SL 03, SIOS Meßtechnik). 

\begin{figure}
	\centering
		\includegraphics[width=0.44\textwidth]{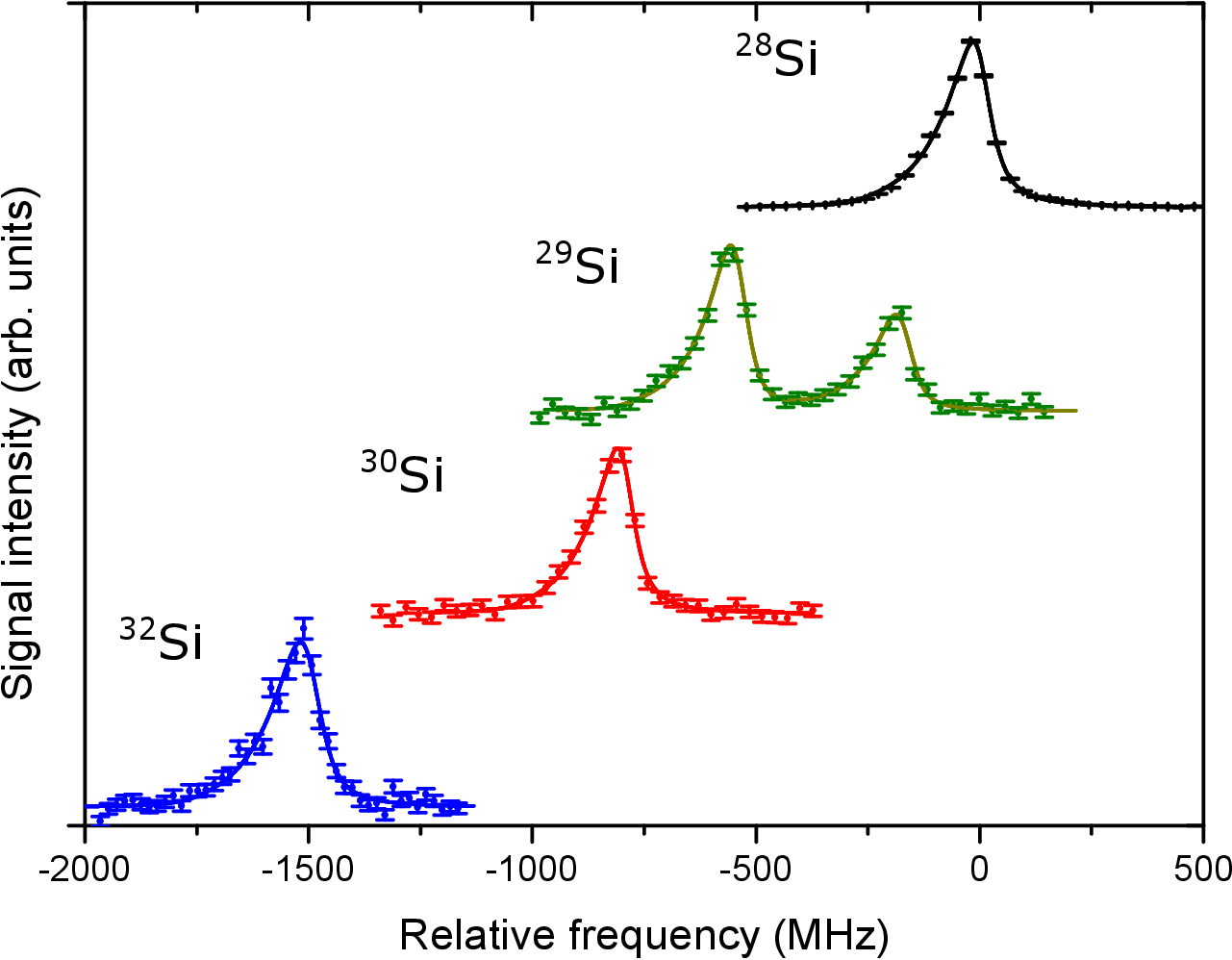}
	\caption{Normalized resonance spectra of $^{28,29,30,32}$Si. Inelastic collisions during the charge exchange process led to slightly asymmetric line shapes and were considered in the fit function \cite{Klose.2012}. The frequency is relative to the centroid of $^{28}$Si.}
	\label{fig:Spectra}
\end{figure}

To avoid optical depopulation along the 70\,cm flight path between CEC and FDR by the strong UV decay, the AOM was used to chop the laser beam so that only unprobed atoms were excited in the FDR.
A continuous ion beam was chosen over a bunched beam to avoid uncertainties caused by a varying temporal overlap between the ion bunch with a time spread of typically $1\,\mu$s and the time width of the laser ($0.3\,\mu$s). More details on the experimental method are presented in the supplementary material.
Examples of the measured spectra are shown in Fig.\,\ref{fig:Spectra}. For the isotope $^{29}$Si with nuclear spin $I=1/2$, the measured hyperfine splitting allowed the determination of the magnetic hyperfine parameter $A_\textnormal{upper}=-252.5(6)$\,MHz. To reduce systematic contributions (e.g., from the wavelength meter), at least five different laser-frequency sets at different beam energies were measured as detailed in \cite{Koenig.2021b}. The results are listed in Tab.\,\ref{tab:IsoShift} together with the literature values of the nuclear charge radii of the stable isotopes \cite{Angeli.2013}. 

The present isotope shifts and absolute radii \cite{Angeli.2013} were used in a King fit procedure \cite{King.1984} to extract the mass and field-shift constants, $K^\mathrm{MS}$ and $F^\mathrm{el}$. The limited amount of reference isotopes, however, restricted the accuracy, particularly for $F^\mathrm{el}$. Hence, additional atomic calculations (method A in Tab.\,\ref{tab:parameters}) were utilized to constrain the field-shift parameter in the King fit. 
The results of the combined determination of $K^\mathrm{MS} = -340.8\,(1.4)$\,GHz\,u and $F^\mathrm{el}=97.0\,(8)$\,MHz/fm$^2$ enable an increased precision in the determination of nuclear charge radii and are used in the further analysis.
Additionally, an independent set of atomic calculations (method B) confirmed the determined atomic parameters (see Table\,\ref{tab:parameters}). Details of the atomic calculations can be found in the supplementary material. \\

\begin{table}
\caption{\label{tab:IsoShift} Measured isotope shifts $\delta\nu^{A,28}=\nu^{A}-\nu^{28}$ relative to $^{28}$Si, differential $\delta \langle r^2 \rangle$, and absolute $R_\mathrm{ch}$ charge radii. The $R_\mathrm{ch}$ of the stable Si isotopes and $^{32}$Ar are taken from \cite{Angeli.2013}. The charge radius of $^{32}$Si was extracted from the isotope shift and the atomic factors given in Tab.\,\ref{tab:parameters}.}
\begin{ruledtabular}
\begin{tabular}{c c c l}
 & $\delta\nu^{A,28}$ &  $\delta \langle r^2 \rangle ^{A,28}$ & $R_\mathrm{ch}$ \\
     &  (MHz)      &  (fm$^2$)       &  (fm) \\
\addlinespace[.2em] \hline \addlinespace[.2em]
$^{28}$Si  & 0                   & 0  	& 3.1224\,(24) \cite{Angeli.2013} \\
$^{29}$Si  & -425.1 (1.1) (2.1)  & -0.030\,(36) & 3.1176\,(52) \cite{Angeli.2013}\\ 
$^{30}$Si  & -805.0 (1.1) (2.1)  & ~0.070\,(29) & 3.1336\,(40) \cite{Angeli.2013}\\ 
$^{32}$Si  & -1505.3 (3.1) (2.1) &  0.195\,(76) & 3.153\,(12) \\ 
$^{32}$Ar  &       -               &  - &   3.3468\,(62) \cite{Angeli.2013}\\
\end{tabular}
\end{ruledtabular}
\end{table}

\begin{figure}
	\centering
		\includegraphics[width=0.44\textwidth]{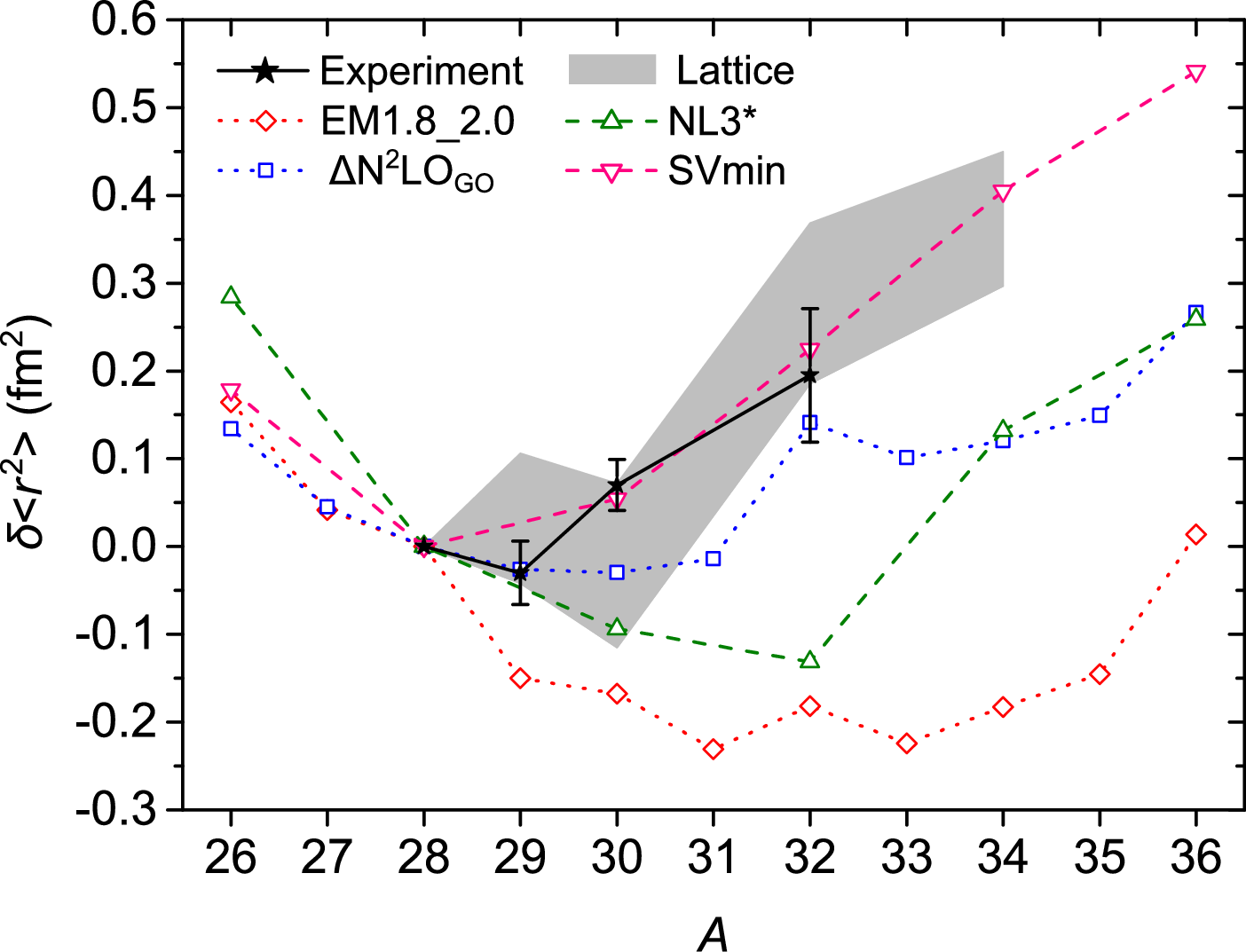}
	\caption{Experimental and theoretical differential mean square charge radii of Si. Only the nuclear lattice calculation provided an uncertainty which is plotted as a gray band. Together with the DFT calculations using the SVmin functional, the lattice results agree with the experimental results.}
	\label{fig:diffChargeRadius}
\end{figure}

\begin{table}
\caption{\label{tab:parameters} Total mass-shift $K^\mathrm{(MS)}$ and field-shift $F^\mathrm{\,(el)}$ parameters from a King fit with the measured Si radii \cite{Angeli.2013}, from two different atomic calculations, and from a combined analysis. In the latter case, the first set of calculations (A) was used to constrain the field-shift parameter in the King-fit procedure. These "combined" $K^\mathrm{(MS)}$ and $F^\mathrm{\,(el)}$ were used in the further analysis and validated by a second independent set of atomic calculations (B).}
\begin{ruledtabular}
\begin{tabular}{ccc}
    Series   &  $K^\mathrm{(MS)}$ & $F^\mathrm{(el)}$   \\ 
     & (GHz~u) & (MHz/fm$^2$) \\
      \colrule
    King fit   &  $-$341.1\,(2.2)          &      114\,(76)               \\ 
    Calculation A          &           &       97.0\,(8)              \\ %-367
    Constrained King fit           &  $-$340.8\,(1.4)   &       97.0\,(8)              \\ 
    Calculation B   &  $-$373\,(24)         &       93.7(3.7)               \\ 
\end{tabular}
\end{ruledtabular}
\end{table}
{\it Nuclear charge radii vs nuclear theory.}
%The theoretical and experimental values are listed in Tab.\,\ref{tab:parameters}, showing a good agreement. The calculations of the field-shift $F^\mathrm{(el)}$ constant yielded small uncertainties, while for the mass-shift constant $K^\mathrm{(MS)}$ larger uncertainties arose that could not be quantified in method B. Contrarily, the experimentally determined uncertainty of the field-shift constant is large and is small for the mass-shift constant. Hence, the weighted mean of the calculated $F^\mathrm{(el)}$ was used to constrain the King fit \textcolor{red}{[we should not used the weighted mean of theoretical values. Perhaps used the two, with different bands?]}, from which the mass-shift constant could then be extracted even more precisely. 
The extracted atomic factors were used to determine the differential mean square charge radius $\delta \langle r^2 \rangle$ of $^{32}$Si, using the expression
\begin{equation}
\delta \langle r^2 \rangle^{A,A'} = \frac{\delta \nu^{A,A'} - \mu^{A,A'} K^\mathrm{(MS)}}{F^\mathrm{(el)}} ,
\label{eq:r2}
\end{equation}
with 
$\mu^{A,A'}=(m_{A}-m_{A'})/((m_{A}+m_e) (m_{A'}+m_e))$, where $m_{A,A'}$ and $m_e$ are the atomic masses of Si and the electron mass, respectively. With the isotope shift given in Tab.\,\ref{tab:IsoShift}, we obtained $\delta \langle r^2 \rangle^{32,28}=0.195\,(76)$\,fm$^2$ resulting in a charge radius of $R(^{32}\textnormal{Si})=3.153\,(12)$\,fm.
Fig.\,\ref{fig:diffChargeRadius} compares the experimental findings with theoretical results from three different complementary many-body methods:
i. Density function calculations (DFT) using two functionals, 
NL3* and SVmin, \cite{Agbemava.2014, Kluepfel.2009}; ii. Valence-Space In-Medium Similarity Renormalization Group (VS-IMSRG) calculations~\cite{PhysRevLett.118.032502,Stro19ARNPS,PhysRevC.102.034320} using two parametrizations of the inter-nucleon interaction, EM$1.8\_2.0$, which generally reproduces ground-state energies well, but underpredicts absolute charge radii~\cite{Simo17SatFinNuc,Stro21Drip}, and $\Delta$N$^2$LO$_{\rm{GO}}$, including explicit $\Delta$-degrees of freedom showing an improved description of radii~\cite{PhysRevC.102.054301}; and iii. nuclear lattice effective field theory calculations \cite{Elhatisari:2022qfr}. Further details of the calculations are included in the supplementary material.

As seen in Fig.\,\ref{fig:diffChargeRadius}, the theoretical results exhibit diverging trends as a function of the neutron number. Within uncertainties, the lattice and the DFT calculation using the SVmin functional show a good agreement with the experiment. 
Interestingly, VS-IMSRG results with the EM$1.8\_2.0$ interaction deviate from the experimental trend in contrast to other regions of the nuclear chart where the same interaction has provided a good description of differential charge radii \cite{deGroote.2020,Ettenauer.2022,Sommer.2022}.  
The calculations with the $\Delta$N$^2$LO$_{\rm{GO}}$(394) interaction are closer to the experimental data, especially with $^{28,29,32}$Si, but fall short in reproducing $^{30}$Si. Furthermore, the VS-IMSRG results predict very different trends beyond $A=32$ compared to lattice and DFT.
A major recent achievement of nuclear lattice calculations has been the description of absolute nuclear charge radii \cite{Elhatisari:2022qfr}. Such absolute results are shown, in Fig.\,\ref{fig:rms_Si} in the supplementary material. 
The VS-IMSRG calculations significantly underestimate the nuclear size, which is an unsolved challenge for most of the {\it ab-initio} calculations of medium and heavy mass nuclei, largely stemming from the input chiral interactions themselves \cite{Gar16,deGroote.2020,Kos21}.
On the other hand, DFT calculations using the SVmin functional overestimate the radii, while the NL3* functional yields good overall radii but misses the experimental trend.\\

{\it Mirror radii and nuclear matter.}
%Despite the diversity and number of experimental studies performed in recent years, the value of the slope of symmetry energy, $L$, could only be constrained to a range between 20 MeV and 120 MeV \cite{Roca-Maza.2018, Horowitz.2014, Vinas.2014}, which is insufficient for precise model predictions. While most of the experimental studies have been performed with stable nuclei, expanding the experimental capabilities to greater isospin-asymmetries with radioactive nuclei is expected to significantly enhance the sensitivity to further constrain $L$ \cite{Piekarewicz.2011, Brown.2017}. So far, only a few experimental techniques suitable for unstable nuclei have been used, such as the measurements of the low-lying E1 strength in $^{130,132}$Sn \cite{Adrich.2005, Klimkiewicz.2007}, $^{68}$Ni \cite{Wieland.2009} and $^{70}$Ni \cite{Wieland.2018}, and the electric-dipole polarizability in $^{68}$Ni \cite{Rossi.2013}. An alternative approach, based on the concept of charge symmetry of the nuclear interaction, proposes to use the difference of charge radii of pairs of mirror nuclei as a proxy for the neutron-skin thickness \cite{Wang.2013, Brown.2017, Yang.2018}, whose correlation and theoretical challenges have been recently discussed in \cite{Reinhard.2022b,Brown.2022,Huang.2023,An.2023}. 
Within the DFT framework, the sensitivity of the mirror charge radii difference, $\Delta R_\textnormal{ch}=R_\textnormal{p}(N,Z)-R_\textnormal{p}(Z,N)$, to $L$, was found to be correlated with $|N - Z| \times L$ \cite{Brown.2017, Roca-Maza.2018, Yang.2018}, where $N$ and $Z$ are the neutron and proton numbers, respectively. This correlation has already been applied to set constraints on $L$ in $^{36}$Ca-$^{36}$S, $^{38}$Ca-$^{38}$Ar \cite{Brown.2020} and $^{54}$Ni-$^{54}$F \cite{Pineda.2021} pairs.
Since the nuclear charge radius of $^{32}$Ar is known \cite{Klein.1996}, our measurement of $^{32}$Si completes the $^{32}$Ar-$^{32}$Si pair with $|N – Z| = 4$. The experimental values are listed in Tab.\,\ref{tab:IsoShift}, and yield $\Delta R_\textnormal{ch} = R_\textnormal{ch}(^{32}\textnormal{Si})-R_\textnormal{ch}(^{32}\textnormal{Ar})=0.194\,(14)$\,fm.

\begin{figure}
	\centering
		\includegraphics[width=0.45\textwidth]{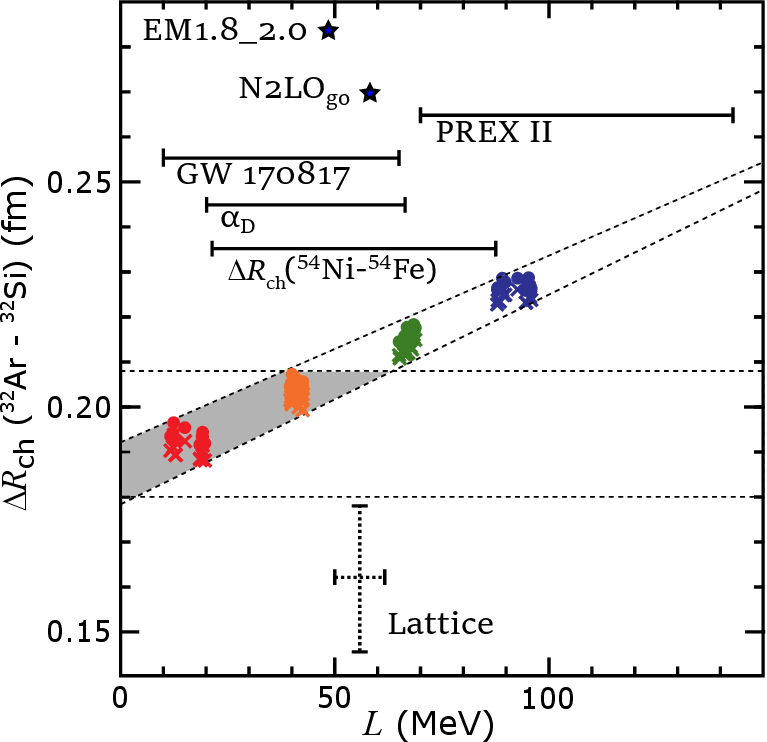}
	\caption{$\Delta R_\textnormal{ch}(^{32}\textnormal{Ar}-^{32}\textnormal{Si}$) as a function of $L$. The experimental $1\sigma$ constraint of $\Delta R_\textnormal{ch}$ is indicated by the horizontal dashed lines. The solid circles are the results of Skyrme EDF, and the crosses are for the CODF calculations. The overlapping area, highlighted in gray, shows our constraint for $L\leq 60$\,MeV.
    It is in good agreement with the result from the $^{54}$Ni-$^{54}$Fe pair \cite{Pineda.2021}, with the findings from the electric dipole polarizability $\alpha_\textnormal{D}$ \cite{Roca-Maza.2015}, the neutron star merger GW 170817 \cite{Raithel.2019} but smaller than the PREX II result \cite{Reed.2021}. Please note that those are only plotted as reference on the $L$ axis and are not correlated to $\Delta R_\textnormal{ch}$. 
    From our theoretical calculations on the lattice and from VS-IMSRG calculations with the EM$1.8\_2.0$ and $\Delta$N$^2$LO$_{\rm{GO}}$ interaction, we deduced $\Delta R_\textnormal{ch}$ and related those with corresponding calculations for $L$ \cite{Elhatisari:2022qfr, Hagen.2016,PhysRevC.102.054301}.} 
    %The results for $L$ agree nicely while $\Delta R_\textnormal{ch}$ disagrees particularly for the VS-IMSRG calculations from the experimental findings.	}
	\label{fig:DeltaR_L}
\end{figure}

To illustrate the correlation between $\Delta R_\textnormal{ch}$ and $L$, DFT calculations were performed for 48 Skyrme energy-density functionals (EDF) \cite{Brown.2013}, and the results are shown in Fig. \ref{fig:DeltaR_L}. The colors in the figure indicate the assumed values for the neutron skin of $^{208}$Pb: 0.12\,fm (red), 0.16\,fm (orange), 0.20\,fm (green), and 0.24\,fm (blue). These calculations are analogous to those carried out for the $A = 36$ mirror pair $^{36}$Ca - $^{36}$S in \cite{Brown.2020}, and are described in the supplementary material.
The correlation and our extracted value of $\Delta R_\textnormal{ch}$ yield a constraint of $L \leq 60$\,MeV. % which is in agreement with previous results obtained from the mirror pairs $^{54}$Ni-$^{54}$Fe, $^{38}$Ca-$^{38}$Ar and $^{36}$Ca-$^{36}$S \cite{Brown.2020, Pineda.2021}.  
As a reference, other experimental constraints of $L$ are shown in the figure without meaning of their $y$-axis position. These constraints come from the Pb neutron-skin thickness (PREX II) \cite{Reed.2021}, the GW170817 binary neutron star merger \cite{Raithel.2019}, the nuclear electric dipole polarizability $\alpha_\textnormal{D}$ \cite{Roca-Maza.2015}, and the $^{54}$Ni$-^{54}$Fe mirror-pair radii \cite{Pineda.2021}. 
Our result agrees well with most of the other findings. However, the PREX II evaluation from Ref.\,\cite{Reed.2021} indicates a stiffer nuclear EOS.
%Furthermore, the present result agrees well with our previous constrains constrains set by the $^{36}$Ca-$^{36}$S, $^{38}$Ca-$^{38}$Ar, and $^{56}$Ni-$^{56}$Fe mirror pairs \cite{Brown.2020, Pineda.2021} whereof only the latter is shown in the figure. 
For comparison, our theoretical results for $\Delta R_\mathrm{ch}$ and $L$ from lattice and VS-IMSRG calculations are depicted in Fig.\,\ref{fig:DeltaR_L}.
%the experimental results are compared with results derived from our lattice calculations. The corresponding $\Delta$R$_\mathrm{ch}$ is plotted in Fig.\,\ref{fig:DeltaR_L}. While the predicted $L$ value agree with the experimentally constrained region, the $\Delta R_\mathrm{ch}$ value is slightly smaller than the experiment.
As VS-IMSRG calculations are not developed yet to calculate properties of nuclear matter, we used our calculated charge radii differences and literature values of $L$ using the $\Delta$N$^2$LO$_{\rm{GO}}$ and EM$1.8\_2.0$ interactions \cite{Hagen.2016,Jia20}. The VS-IMSRG calculations overestimate $\Delta R_\mathrm{ch}$, while the lattice calculations yield a slightly smaller value, mainly due to an underestimation of $R_\mathrm{ch}(^{32}$Ar) compared to experimental data. The results for $L$, however, are in good agreement with complementary calculations available in the literature, such as Quantum  Monte Carlo  \cite{Steiner.2012, Gandolfi.2013}, energy density functionals \cite{Reinhard.2021,Reinhard.2022}, chiral effective field theory calculations \cite{Drischler.2020}, and a combined analysis of astrophysical data with PREX II and chiral effective field theory \cite{Essick.2021}.
All of those theoretical results agree with our experimental constraint of $L \leq 60$ MeV.\\

{\it Conclusions and Outlook}.
We performed collinear laser spectroscopy of $^{28,29,30,32}$Si and determined the nuclear charge radius of $^{32}$Si. Our experimental result provides an essential benchmark for the development of theoretical models and is in good agreement with {\it ab initio} lattice predictions and DFT calculations using the SVmin functional. In contrast to the results for different regions of the nuclear chart \cite{Kos21,deGroote.2020}, VS-IMSRG calculations fall short of reproducing the charge radii of silicon isotopes. 
Beyond $A=32$, the applied theoretical models significantly deviate in their trends, motivating further research of neutron-rich silicon isotopes. 

The present radius of $^{32}$Si, combined with the literature value of $^{32}$Ar, allowed testing the correlation between the nuclear mirror radii differences and the slope of the symmetry energy of the equation of state of the nuclear matter. The result suggests a value of $L \leq 60$\,MeV, that is in good agreement with constraints obtained from other mirror pairs \cite{Brown.2020,Pineda.2021} and different experimental observables, such as gravitational waves of the binary neutron star merger \cite{Abbot.2017} and nuclear reactions \cite{Tsang.2012}. 

The silicon isotopic chain further exhibits unique features that make these nuclear systems particularly challenging and attractive for our understanding of the nuclear many-body problem. Of special future interest is the study of the suggested “doubly-magic” nuclei $^{34}$Si \cite{Baumann.1989} and $^{42}$Si \cite{Fridmann.2005,Bastin.2007}, as well as the suggested “bubble” structure in $^{34}$Si \cite{Yao.2013,Mutschler.2016, Duguet.2017}.  
Moreover, a future charge radius measurement of the neutron-deficient isotope $^{22}$Si could be combined with its mirror $^{22}$O to form their mirror radii difference with the largest proton-neutron asymmetry $|N - Z| = 6$ of all reasonably accessible pairs, thus resulting in the highest sensitivity to $L$.
%Both neutron-deficient and neutron-rich isotopes will be reachable at  FRIB.

%The lower production rate of those isotopes will require a more efficient detection, which will become possible with RISE at BECOLA, where collinear resonance ionization spectroscopy is applied. Here, excitations with deep UV-light from the highly populated atomic ground state with subsequent photoionization are possible, which shall boost the efficiency by two orders of magnitude.

\begin{acknowledgements}
This work was supported in part by the National Science Foundation under Grant No. PHY-21-11185 and PHY-21-10365, and the U.S. Department of Energy under the grants DE-SC0021176, DE-SC0021152, DE-SC0013365, DE-SC0023658, SciDAC-5 NUCLEI Collaboration. Calculations of isotope shift constants have been supported by the Russian Science Foundation Grant No. 19-72-10019. We thank the Center for Information Technology of the University of Groningen for their support and for providing access to the Peregrine high-performance computing cluster. JK acknowledges the support of a Feodor Lynen Fellowship of the Alexander-von-Humboldt Foundation. {YZM was supported by the National Natural Science Foundation of China under Grants No. 12105106 and China Postdoctoral Science Foundation under Grant No. BX20200136. The work of UGM  and SE was supported in part by the European Research Council (ERC) under the European Union’s Horizon 2020 research and innovation program (grant agreement No. 101018170). The work of UGM was further supported by VolkswagenStiftung (Grant no. 93562) and by the CAS President’s International Fellowship Initiative (PIFI) (Grant No. 2018DM0034). For the lattice calculations, we acknowledge computational resources provided by the Oak Ridge Leadership Computing Facility through the INCITE award ``Ab-initio nuclear structure and nuclear reactions'', the Southern Nuclear Science Computing Center, the Gauss Centre for Supercomputing e.V. (www.gauss-centre.eu) for computing time on the GCS Supercomputer JUWELS at the J{\"u}lich Supercomputing Centre (JSC), and the Institute for Cyber-Enabled Research at Michigan State University. VS-IMSRG calculations are supported by NSERC under grants SAPIN-2018-00027 and RGPAS-2018-522453 as well as the Arthur B. McDonald Canadian Astroparticle Physics Research Institute. We thank S. R. Stroberg for the imsrg++ code~\cite{Stro17imsrg++} used to perform these calculations. Computations were performed with an allocation of computing resources on Cedar at WestGrid and the Digital Research Alliance of Canada.}
\end{acknowledgements}

\bibliographystyle{aipnum4-1}
\bibliography{literature}

\newpage % Ende vom Text
\pagestyle{empty}
%\null\newpage

\nocite{Anton.1978,Klose.2012,DIRAC19,Saue:2020,Dyall:2016,Landau2001,PhysRevA.102.052812,Oleynichenko_EXPT,Oleynichenko:20,Dyall:2016,PhysRevA.102.052812,shabaev1985mass,palmer1987reformulation, shabaev1988nucl,shabaev1994relativistic,Penyazkov:2023,Oleynichenko:20,Dyall:2016,DIRAC19,Saue:2020,Oleynichenko_EXPT,Oleynichenko:20,dzuba96pra,kahl19cpc,Elhatisari:2022qfr,Ma:2023ahg,PhysRevC.102.034320,PhysRevC.105.014302,PhysRevC.83.031301,Simo17SatFinNuc,PhysRevC.102.054301,SHIMIZU2019372,Brown.2006,Brown.2020,Brown.2017,Brown.1998,Pineda.2021}
\newpage
\section*{Supplementary Material}

\begin{figure}
	\centering
		\includegraphics[width=0.460\textwidth]{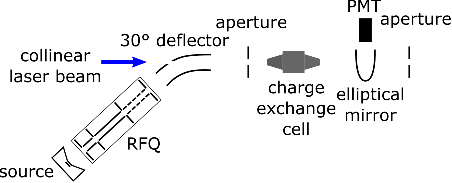}
	\caption{Schematic of the BECOLA beamline. In the radio-frequency-quadrupole trap (RFQ) the beam was cooled and extracted as a continuous beam. Laser and ion beams were superimposed and aligned through two 3-mm apertures at 2.1\,m distance. The beam was neutralized through charge-exchange reactions with Na vapor. Fluorescence light was collected by an elliptical mirror set and guided to a photo-multiplier tube (PMT) outside the vacuum chamber. Further ion optics for beam deflection and collimation are not shown.
	}
	\label{fig:setup}
\end{figure}

{\it Methods.} A schematic of the BECOLA setup is depicted in Fig.\,\ref{fig:setup}. First measurements were performed with stable Si beam from the PIG source in order to optimize the system in terms of efficiency, signal-to-noise ratio (SNR) and line shape. 
Due to the branching ratio of the upper state, 94\,\% of the atoms are lost into a dark state after one excitation. Since the time of flight was $\approx 1.6\,\mu$s to pass the 70-cm distance between CEC and FDR, in which already laser-ion interactions were taking place, higher laser intensities led to a decreasing SNR and to a distorted line shape.
In order to overcome this, the AOM was used to chop the laser beam. By blocking the laser beam for $1.6\,\mu$s, there were always unprobed atoms in the FDR. Then, the beam was unblocked for $0.3\,\mu$s, which corresponds to the time of flight through the FDR. Since the dark pumping was strongly reduced, a relatively high laser power of 4\,mW with a laser beam diameter of 2\,mm was used, which yielded the highest SNR. Compared to lower laser powers, the impact on the line shape was marginal. Nevertheless, inelastic atom-atom collisions during the charge-exchange process caused a tail to slower atom velocities \cite{Anton.1978,Klose.2012} as can be seen in Fig.\,\ref{fig:Spectra}.
An electronic coincidence unit was used to accept only PMT signals while the AOM was on to avoid background counts during the AOM-off times, e.g., from the continuous ion beam. 
The continuous beam was preferred over a bunched beam to avoid uncertainties caused by a varying temporal overlap between the ion bunch with a time width of $1\,\mu$s and the laser ($0.3\,\mu$s).\\

{\it Atomic calculations.}
In method A, the field-shift constant was calculated using the Fock-Space coupled cluster method with single and double excitation amplitudes in the {\sc dirac} code~\cite{DIRAC19,Saue:2020}. The augmented ACV4Z~\cite{Dyall:2016} basis set, with two additional layers of diffuse functions added in an even tempered fashion, was employed and all electrons and virtual orbitals were included in the correlating space. The model space consisted of the 3p 4s (3d 4p 5s 5p 4f 4d) orbitals, where the orbitals in parentheses are in the intermediate Hamiltonian model space~\cite{Landau2001}. The uncertainty is estimated using the same approach as in~\cite{PhysRevA.102.052812}. In addition we include the effect of full triple cluster amplitudes by employing the {\sc exp-t} code~\cite{Oleynichenko_EXPT,Oleynichenko:20} within the V4Z~\cite{Dyall:2016} basis set, keeping the 1s electrons frozen and setting the virtual cutoff at 5 a.u. to include the 4-7s 3-7p 3-5d 4-5f 6g orbitals in the active virtual space. 
The approaches of method A are analogous to the one used for the isotope shift constants of Sn in~\cite{PhysRevA.102.052812} and we refer there for more details on the methodology.

In method B, we have used expressions for the relativistic nuclear recoil Hamiltonian~\cite{shabaev1985mass,palmer1987reformulation, shabaev1988nucl,shabaev1994relativistic} to calculate the normal and specific mass shift constants. Corresponding matrix elements were calculated within the code developed in Ref.~\cite{Penyazkov:2023}. The field shift constant $F$ has been defined as $F=d\nu/d\left\langle r^2 \right\rangle$. Electronic structure calculations of isotopic shift constants were performed using the relativistic Fock-Space coupled cluster method with single, double, and triple excitation amplitudes (FS-CCSDT)~\cite{Oleynichenko:20}. The calculations were carried out using the manually extended AE3Z~\cite{Dyall:2016} basis set, which included the addition of 5$s$-, 3$p$-, 2$d$-, and 1$f$-type diffuse functions to the original AE3Z~\cite{Dyall:2016} basis set. The Dirac-Coulomb Hamiltonian was employed, and the calculations were performed using the {\sc dirac}~\cite{DIRAC19,Saue:2020} and {\sc exp-t}~\cite{Oleynichenko_EXPT,Oleynichenko:20} codes. A further correction on the extended basis set was calculated using the Fock-Space coupled cluster method with single and double excitation amplitudes (FS-CCSD) method. In these calculations, the basis set was expanded to the manually extended AAE4Z~\cite{Dyall:2016} basis set. This expansion included the addition of 7$s$-, 7$p$-, 5$d$-, and 6$f$-type diffuse functions, as well as 2$p$-, 4$d$-, 3$f$-type tight functions in an even tempered fashion, and 4$h$-type functions. In these calculations all electrons were included in correlation treatment and the virtual energy cutoff was set to 500 a.u. In method B we considered two sources of theoretical uncertainty: (i) basis set incompleteness, which was conservatively estimated by comparing results obtained with the extended AAE4Z and extended AE3Z basis sets within the FS-CCSD method, and (ii) uncertainty arising from the treatment of electronic correlation effects incompletely. This was estimated by calculating the difference between the results obtained using the FS-CCSDT and FS-CCSD methods within the AE3Z basis set. The total uncertainty for each atomic factor was calculated as the square root of the sum of the squares of these two uncertainties. The resulting values of the normal and specific mass shift atomic constants are $418(17)$ GHz~u and $-791(17)$ GHz~u, respectively.

The mass-shift factors were also calculated within a complementary approach using of a combination of configuration interaction and many-body perturbation theory (CI+MBPT)~\cite{dzuba96pra}, using the AMBiT code~\cite{kahl19cpc}. The CI calculation included all single and double excitations from the reference configurations ($3p^2$, $3p4s$) up to 20spdfg, while hole excitations from the 3s orbital were included. The MBPT basis included all virtual orbitals up to 40spdfgh and one-, two- and three-body MBPT diagrams involving the reference configurations were included. The obtained result of $-367$ GHz~u for the total mass shift is in good agreement with the values listed in Tab.\,\ref{tab:parameters}, providing an independent confirmation of these results. \\

{\it Nuclear lattice effective field theory calculations.}
Our {\it ab initio} lattice results for $L$ were determined from the calculations of pure neutron matter in Ref.~\cite{Elhatisari:2022qfr}.  The lattice simulations for the charge radii are new calculations based upon the N3LO chiral interactions described in Ref.~\cite{Elhatisari:2022qfr} with two additional improvements made.  Rather than a global fit to all nuclei, we fit the three-nucleon coefficients $c_E^{(l)}$ and $c_E^{(t)}$ to ensure good agreement with the binding energies of the silicon isotopic chain.  We also use the rank-one operator method introduced in Ref.~\cite{Ma:2023ahg} to compute the charge radii. \\

{\it VS-IMSRG calculations.}
Valence-Space In-Medium Similarity Renormalization Group calculations \cite{PhysRevC.102.034320} were performed by working in a 13 major-shell harmonic oscillator (HO) space with an additional cut $E_{\rm 3max}$=24 truncation on storage of three-nucleon (3N) matrix elements \cite{PhysRevC.105.014302}. 
The two plus three-nucleon EM1.8\_2.0 \cite{PhysRevC.83.031301,Simo17SatFinNuc} and $\Delta$N$^2$LO$_{\rm GO}$ \cite{PhysRevC.102.054301} interactions were employed throughout the study. We decoupled a proton $sd$ and neutron $sdf_{7/2}p_{3/2}$ multishell valence space Hamiltonian above $^{16}$O core for $^{26-36}$Si. The final exact diagonalization was performed using the \textsc{KShell} shell-model code \cite{SHIMIZU2019372}.\\

\begin{figure}
	\centering
		\includegraphics[width=0.460\textwidth]{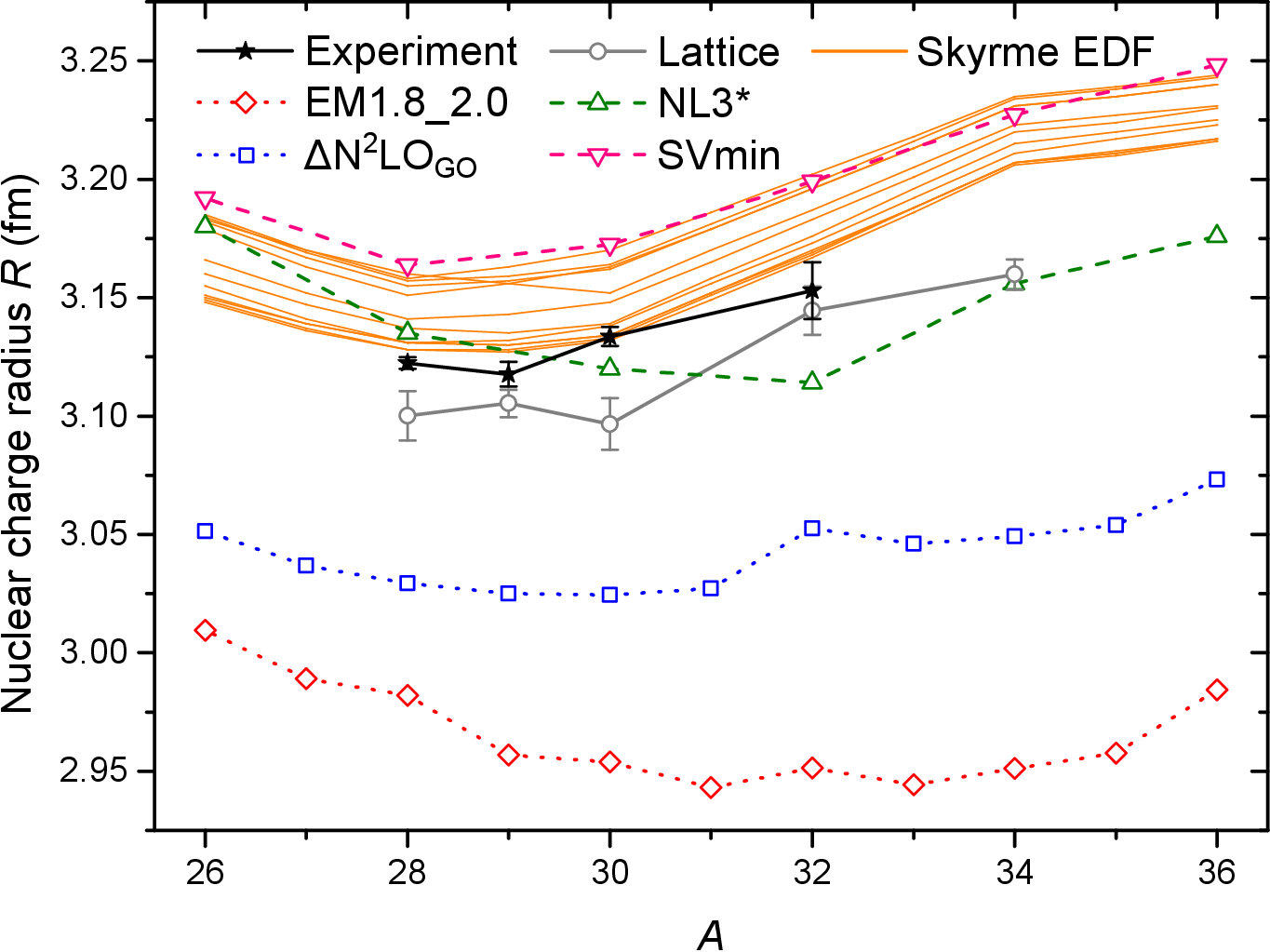}
	\caption{Experimentally and theoretically determined root-mean square charge radii $R$ of Si. %Contrary to DFT and VS-IMSRG, the calculation on the lattice shows a good agreement to the experimental results in absolute size and trend.
 The band of orange lines show the results for $R$ obtained with the 12 Skyrme EDF points colored orange in \ref{fig:DeltaR_L} that correspond to an assumed $^{208}$Pb neutron skin of 0.16\,fm.
}
	\label{fig:rms_Si}
\end{figure}

{\it EDF and CODF calculations}. The orbital occupation numbers for the spherical EDF calculations were constrained to those calculated in the $sd$ model space with the USDB Hamiltonian \cite{Brown.2006}. The orbital occupations for $^{32}$Si are given in Table \ref{tab:occupation}. The occupation numbers for $^{32}$Ar are the same as those for $^{32}$Si but with the protons and neutrons interchanged.

These calculations are analogous to those carried out for the $A = 36$ mirror pair $^{36}$Ca - $^{36}$S in \cite{Brown.2020}. For $A = 36$ \cite{Brown.2020} it was found that the Skyrme results for $\Delta R_\textnormal{ch}$ were shifted compared to those obtained with covariant density-functional (CODF) calculations. This difference was traced to the difference in the form of the spin-orbit potential between EDF and CODF. The Skyrme EDF spin-orbit potential as used in \cite{Brown.2017} and in Fig.\,\ref{fig:DeltaR_L}, has both isoscalar (IS) and isovector (IV) $p$-wave components. In contrast, the CODF spin-orbit potential is purely IS. When the Skyrme spin-orbit form was changed to purely IS form as used, for example, in \cite{Brown.1998}, the EDF and CODF results agreed rather well for $A = 36$. The Skyrme results for $A = 32$ obtained with a purely IS potential are shown by the colored crosses in Fig.\,\ref{fig:DeltaR_L}. In contrast to $A = 36$, for $A = 32$ the IS$+$IV and IS results are similar, as they were for $A = 54$ \cite{Pineda.2021}.

The analysis for the $A = 54$ pair $^{54}$Ni - $^{54}$Fe in \cite{Pineda.2021} included a correction to the spherical EDF calculations for the quadrupole deformation. The correction was obtained from the data for the B(E2) 0$^{ + }_{1}$ to 2$^{ + }_{1}$ transition in $^{54}$Fe, together with shell-model calculations for both $^{54}$Fe and $^{54}$Ni.
For the $A = 32$ pair, the $B(E_2)(0^{+}_{1} \rightarrow 2^{+}_{1})$ needed for the deformation correction can be obtained from experiment. Then the corrections were calculated using the formulation in \cite{Pineda.2021}: For $^{32}$Ar, $B(E_2)$ = 54(14) e$^{2}$ fm$^{4}$ and $\delta$R$_{p}$($^{32}$Ar) = 0.0069 fm, and for $^{32}$Si, B(E2) = 26(7) and $\delta$R$_{p}$($^{32}$Si)  = 0.0055(14) fm. The total correction for $A=32$ is $\delta$R$_{p}$($^{32}$Ar) - $\delta$R$_{p}$($^{32}$Si) = 0.0014(22)\,fm. This is small compared to the experimental value of $\Delta$R$_\mathrm{ch}$ = 0.194(14) fm.

\begin{table}[b]
\caption{The sd shell orbital occupation numbers for $^{32}$Si obtained with the USDB Hamiltonian.}
\label{tab:occupation}
\begin{ruledtabular}
\begin{tabular}{c c l}
orbital & proton & neutron\\
\hline
0$d_{3/2}$ & 0.22 & 2.55\\
1$s_{1/2}$ & 0.43 & 1.61\\
0$d_{5/2}$ & 5.35 & 5.84\\
\end{tabular}
\end{ruledtabular}
\end{table}

\end{document}